\documentstyle[12pt,a4,epsf]{article}
\begin{document}
\input psfig
\bibliographystyle{unsrt}
\baselineskip= 18pt
\pagenumbering{arabic}
\pagestyle{plain}
\def\bit{\begin{itemize}}
\def\eit{\end{itemize}}
\def\half{{1\over 2}}
\def\OO{\Omega}
 \def\aa{\alpha}
 \def\bb{\beta}
 \def\bk{{\bf k}}
 \def\bkp{{\bf k'}}
 \def\bqp{{\bf q'}}
 \def\bq {{\bf q}}
 \def\EE{\Bbb E}
 \def\EEx{\Bbb E^x}
 \def\EEo{\Bbb E^0}
 \def\LL{\Lambda}
 \def\PP{\Bbb P^o}
 \def\RR{{\bf R}}
 \def\rr{\rho}
 \def\SS{\Sigma}
 \def\ss{\sigma}
 \def\ll{\lambda}
 \def\dd{\delta}
 \def\ww{\omega}
 \def\ll{\lambda}
 \def\DD{\Delta}
 \def\GG{\Gamma}
 \def\DDt{\tilde {\Delta}}
 \def\kr{\kappa\lb \LL\rb}
 \def\PPx{\Bbb P^{x}}
 \def\gg{\gamma}
 \def\kk{\kappa}
 \def\tt{\theta}
 \def\eps{\epsilon}
 \def\lb{\left(}

 \def\rb{\right)}
 \def\prt{\tilde p}
\def\pt{\tilde {\phi}}
 \def\nab2{\nabla^2}
 \def\hal{{1\over 2}\nabla ^2}
 \def\bg{{\bf g}}
 \def\bx{{\bf x}}
 \def\bu{{\bf u}}
 \def\bv{{\bf v}}
 \def\by{{\bf y}}
 \def\hag{{1\over 2}\nabla}
 \def\beq{\begin{equation}}
 \def\eeq{\end{equation}}
 \def\bea{\begin{array}}
 \def\eea{\end{array}}
 \def\cosech{\hbox{cosech}}
 \def\fr{\frac}
 \def\Tr{{\mbox{Tr}}}
 \def\d{\partial}
 \def\la{\langle}
 \def\ra{\rangle}
 \def\n{\hfil\break}
 \def\rw{\rule[-3mm]{0mm}{8mm}}
\title{Lattice String Breaking and Heavy Meson Decays}
\author{I T Drummond and R R Horgan \\
        Department of Applied Mathematics and Theoretical Physics \\
        University of Cambridge \\
        Silver St \\
        Cambridge, England CB3 9EW}
\maketitle
\begin{abstract}
We show how string breaking on the lattice, treated as a mixing effect, can be
related to decay rates for heavy quark systems.  We use this to make a preliminary calculation 
of the energy split at maximum mixing for static quarks in QCD from
the decay rate for $\Upsilon(4S)\rightarrow B{\bar{B}}$~.
We extend the calculation to achieve rough estimates for the contributions of 
channels involving $B$, $B^*$, $B_s$ and $B_s^*$ mesons to the width of
the $\Upsilon(5S)$~. 

\end{abstract}
\vfill
DAMTP-98-127
\pagebreak

\section{Introduction}

It has been recognised for a long time that, just as the confining string-like
interquark potential provides a good basis for understanding valence quark
dynamics, so the breaking of the string provides a natural way of understanding
the influence of sea quarks on decay processes \cite{Isgur1,Isgur2}. 
Lattice measurements of the confining QCD potential, in quenched and unquenched 
calculations, are well understood \cite{UKQCD,BaSch1,BaSch2,BaSch3,CPPACS,Burk,Ka}. Moreover 
the simplest type of quark model using this potential gives a reasonable account 
of the observed heavy meson spectrum. Recently there has been progress in the 
attempt to observe string breaking effects directly on the lattice \cite{PhWi,KnSo,DeT}.

In this paper we investigate how sufficiently accurate lattice measurements of 
the interquark potential and string breaking effects might be combined to implement 
a programme of computation of decay rates for heavy quark systems. The analysis 
depends on a two-channel mixing picture of the string breaking process suggested 
by a crude model proposed previously \cite{ITD1,ITD2,ITD3}. The basis of the 
mechanism is the fact that the light quark-anti-quark pair is produced (or absorbed) 
in a triplet spin state. Although our picture of string breaking is different in detail from
that of Isgur and Paton \cite{Isgur1} the results are still consistent with the 
phenomenological $~^3P_0$-model~. 

\section{Crude Model}

The crude model described previously
\cite{ITD1,ITD2,ITD3}, was designed to provide a simple two-channel
mixing model of string breaking. It suggests that when the string breaks 
each light quark and anti-quark materialises in the neighbourhood of the 
appropriate static anti-quark or quark. In this respect it differs
from the string breaking model of Isgur and Paton \cite{Isgur1} who viewed the
materialisation of the light quark-anti-quark pair as a local event on the string.
As a justification of our picture we point out that to a first approximation 
the local gauge field energy density does not change as the static quarks are separated. 
There is no real mechanism therefore for the local production of the light pair.
The breaking event is only possible when the total energy distributed along
the string is sufficient to support light pair creation. Our model then 
suggests that this creation occurs essentially instantaneously in such a way
that the gauge field distribution supported by the static quarks is 
cancelled by the opposing gauge field structure of the light quarks.
It is this idea of instantaneous transition that we exploit below in our development
of the model. However the outcome of the these considerations is not
in contradiction with the $~^3P_0$ model that is supported by the analysis of 
Isgur and Paton, at least for the decay processes to which we apply our ideas.

As was pointed out in the original description of the crude model \cite{ITD1,ITD2,ITD3},
the dynamical quark-anti-quark pair are produced in a correlated spin state. 
Here we examine this mechanism a little more closely. In the model the
strong coupling graph that corresponds to the production of the pair is
shown in Fig 1~. The quark or anti-quark binds in an $S$-wave with the appropriate
static anti-quark or quark. Choosing 0 as the (imaginary) time direction and 1 as the 
direction of spatial separation of the static quarks, we see that the 
rules for quark propagation \cite{ITD1,ITD2,ITD3,CREUTZ,MonMun} yield an amplitude 
with spin structure of the form
\beq
\fr{1+\gg_0}{2}\fr{1+\gg_1}{2}\fr{1-\gg_0}{2}
                                 =\fr{1}{2}\left(\bea{cc}0&\ss_1\\0&0\eea\right)~~.
\eeq
This can be interpreted as meaning that the light quark-anti-quark pair are produced
with a spin wave function 
\beq
\xi_{\aa\bb}=\fr{1}{\sqrt{2}}(\ss_i{\hat{\RR}}_i)_{\aa\bb}~~,
\label{LQWF1}
\eeq
where $\aa$ and $\bb$ are the spin labels for the quark and anti-quark respectively
and $\hat{\RR}$ is the unit vector pointing along the line of separation of the
the static quarks. This shows that the light quark-anti-quark pair are produced
in a triplet state with zero spin component along the line of separation of the static quarks.


\section{Basic Mixing Scenario}

In order to explain the basic idea we first consider a model
in which the particles of the theory have no spin but do
experience confinement. The non-abelian Higgs models are of this type \cite{PhWi,KnSo}.
Moreover they exhibit string breaking on the lattice of a type consistent 
with a two-channel mixing scenario \cite{ITD1,ITD2,ITD3}. The channels involved are 
\bit
\item[1)] a static quark-anti-quark $\{Q{\bar Q}\}$ system with a separation $R$
and  with a connecting flux string
\item[2)] a heavy-light meson-meson system $\{Q{\bar q}, q{\bar Q}\}$~. 
\eit
For the moment, we ignore the spins of the quarks and treat them as scalar particles.

On the lattice it is possible to measure the string-string correlator
${\cal G}_{SS}(T)$ as the $R\times T$ Wilson loop. It has the form
\beq
{\cal G}_{SS}(T)\sim e^{-V_{Q{\bar Q}}(R)T}~~,
\eeq
where $V_{Q{\bar Q}}(R)$ can be interpreted as the interquark potential experienced by
the static quarks. Lattice measurents are consistent with the Cornell potential 
\cite{CORN1,CORN2,CORN3}
\beq
V_{Q{\bar Q}}(R)=C-\fr{\aa}{R}+\ss R~~,
\label{CORNELL}
\eeq 
where $\ss$ is the string tension.
The correlator for two-meson states over the time interval $T$ can be obtained
from measurements of observables appropriate to the two-meson state. Its
behaviour is of the form
\beq
{\cal G}_{MM}(T)\sim e^{-E_{MM}(R)T}~~,
\eeq
where $E_{MM}(R)=2E_M+V_{MM}(R)$ and $E_M$ is the energy of a single static meson 
and $V_{MM}(R)$ is the potential interaction between the two static mesons. This 
potential energy has not been measured in QCD but does show up in the Higgs model 
calculations.  We might expect it to have a form like \cite{ITD3}
\beq
V_{MM}(R)=W_Me^{-mR}~~,
\label{TRANSV}
\eeq
where $m$ is the mass of a light meson that can be exchanged between
the heavy mesons and $W_M$ is an energy determining the overall strength
of the interaction.  

This description is of course an oversimplification in the presence of dynamical
matter fields. This is clear from the Higgs model calculations \cite{PhWi,KnSo} and in QCD
from the thermal Polyakov loop calculations \cite{DeT}. The reason is that string 
breaking can occur as a process that leads to a mixing of the string and two-meson states.
Consistently with these static model calculations we may represent the mixing process 
by a transition potential $V_{I}(R)$~. On the basis of the crude model we might expect 
it to have the form \cite{ITD1,ITD2,ITD3}
\beq
V_{I}(R)=We^{-m_qR}~~,
\label{TRANSPOT}
\eeq
where $m_q$ is the mass of the dynamical matter field and $W$ is the energy
parameter determining the strength of the transition potential. It follows that 
we are dealing with a two-channel interaction matrix of the form
\beq
V(R)=\left(\bea{cc}V_{Q{\bar Q}}(R)&V_{I}(R)\\
                     V_{I}(R)&E_{MM}(R)\eea\right)~~.
\eeq
The energies measured directly on the lattice are the eigenvalues of $V(R)$~.
They are
\beq
V_{\pm}(R)=\fr{1}{2}\left\{V_{Q{\bar Q}}(R)+E_{MM}(R)
             \mp\sqrt{(V_{Q{\bar Q}}(R)-E_{MM}(R))^2+4(V_{I}(R))^2}\right\}~~.
\label{EIGV}
\eeq
We keep the convention used previously that $V_{+}(R)$ is the lower of the
two eigenenergies. If $V_{I}(R)$ is sufficiently small then the eigenenergies 
are dominated by the diagonal elements of $V(R)$~. We can define a critical value
of $R_c$ for which these diagonal values are equal
\beq
V_{Q{\bar Q}}(R_c)=E_{MM}(R_c)~~.
\eeq
When $R=R_c$ the the split in the eigenenergies is proportional to $V_{I}(R_c)$, we have
\beq
V_{\pm}(R_c)=V_{Q{\bar Q}}(R_c)\mp V_{I}(R_c)~~.
\eeq
For $R<<R_c$ we expect $V_{Q{\bar Q}}(R)<<E_{MM}(R)$ with the result
$V_{+}(R)\simeq V_{Q{\bar Q}}(R)$ and $V_{-}(R)\simeq E_{MM}(R)$~.
For $R>>R_c$ we expect $V_{Q{\bar Q}}(R)>>E_{MM}(R)$ with the result
$V_{-}(R)\simeq V_{Q{\bar Q}}(R)$ and $V_{+}(R)\simeq E_{MM}(R)$~.
The interchange that occurs near $R=R_c$ is the mixing phenomenon.
Just as in refs \cite{ITD1,ITD2,ITD3} we can compute a mixing angle $\tt$ given by
\beq
\tan\tt=\fr{-(V_{Q{\bar Q}}(R)-E_{MM}(R))
            +\sqrt{(V_{Q{\bar Q}}(R)-E_{MM}(R))^2+4(V_{I}(R))^2}}{2V_{I}(R)}~~.
\eeq
We have
\beq
V(R)=O\left(\bea{cc}V_{+}(R)&0\\
                    0&V_{-}(R)\eea\right)O^{-1}~~,
\eeq
where
\beq
O=\left(\bea{cc}\cos\tt&-\sin\tt\\
                    \sin\tt&\cos\tt\eea\right)
\label{ORTHO}
\eeq
As $R$ passes from below to above $R_c$, the mixing angle increases from 0 to $\pi/2$
hitting $\pi/4$ when $R=R_c$~. The goal then is to use lattice measurements
to determine the eigenenergies and the $R$-dependence of the mixing angle.
By those means we can determine the full two-channel interaction matrix, $V(R)$~.
We discuss this in more detail below.

Once we know the elements of $V(R)$ we can make
the two-channel equivalent of the Born-Oppenheimer approximation and
replace the static quarks by heavy non-relativistic quarks and mesons.
The effective Schr\"odinger equation is
\beq
i\fr{\d}{\d t}\psi(\RR)=\left(\bea{cc}-\fr{1}{2\mu_Q}\nab2&0\\
                        0&-\fr{1}{2\mu_M}\nab2\eea\right)\psi(\RR)+V(R)\psi(\RR)~~,
\label{SE}
\eeq
where 
\beq
\psi(\RR)=\left(\bea{c}\psi_Q(\RR)\\\psi_M(\RR)\eea\right)~~,
\eeq
and $\mu_Q$ and $\mu_M$ are the reduced masses for the heavy quarks and heavy 
mesons in their respective channels. These also can be obtained from appropriate
lattice measurements of heavy single meson states. 

We can exploit eq(\ref{SE}) by first ignoring the off-diagonal elements of $V(R)$~.
We then calculate the heavy $Q{\bar Q}$ bound states with energies $E_{Q{\bar Q}}$~.
If this energy lies above the two-meson channel threshold then we can compute the
decay rate as
\beq
\GG=2\pi\rho(k)|T(k)|^2~~,
\label{GR1}
\eeq
where $E_{Q{\bar Q}}$ is the energy of the bound state $Q{\bar Q}$ system in the potential
$V_{Q{\bar Q}}(R)$~, $\bk$ is the relative momentum of the final state mesons, 
$\rho(k)$ is the density of states factor. The energy of the two-meson state is 
$E(k)$ which satisfies
\beq
E(k)=\fr{1}{2\mu_M}k^2+2E_M=E_{Q{\bar Q}}~~.
\eeq
The transition amplitude is
\beq
T(k)=\la\bk|V_{I}|\psi_Q\ra~~,
\eeq
$|\bk\ra$ being the wavefunction, with incoming scattered wave, for the two meson system in
the presence of the potential $V_{MM}(R)$~. As a first approximation we could neglect
the the final state interactions represented by the scattered part of the the wave 
function and replace $|\bk\ra$ with the plane wave $e^{i\bk.\RR}$~. We have then
\beq
T(k)=\int d^3\RR e^{-i\bk.\RR}V_{I}(R)\psi_Q(R)
=4\pi\int_{0}^{\infty}dRR^2\left(\fr{\sin kR}{kR}\right)V_{I}(R)\psi_Q(R)~~.
\label{GR2}
\eeq
In the above it is assumed that the static approximation for the mesons is accurate
in the sense that
\beq
E_{Q{\bar Q}}-2E_M=M_{Q{\bar Q}}-2M_M~~.
\label{EGAP}
\eeq
where $M_{Q{\bar Q}}$ and $M_M$ are the actual masses of the heavy quark state
and the heavy-light meson state respectively. To the extent that the results
are only approximate the left side of eq(\ref{EGAP}) should be replaced
by the right side in an actual estimate of the decay rate.

\section{Lattice Measurement of the Interaction Matrix}

On the lattice we can measure $V_{\pm}(R)$ \cite{PhWi,KnSo,DeT}. If we are confident of 
our parametrizations of the elements of $V(R)$, then it is likely that we can 
determine them by performing a fit to the eigenenergies provided they are known
sufficiently accurately and on a sufficiently fine spatial lattice. 

In principle, the the eigenenergies are not sufficient to fix the
elements of $V(R)$~. A superior approach, therefore, would be to measure 
also the mixing angle $\tt$~. The results obtained in the $SU(2)$-Higgs 
models for the $R$-dependence of the ground state overlap of the lower 
eigenmode suggest that this may well be possible \cite{PhWi,KnSo}~. The two-channel model 
suggests that the contribution of the lower energy eigenmode to the Wilson loop is
\beq
{\cal G}_{SS}(R)\simeq \GG\cos\theta e^{-V_+(R)T}\GG\cos\theta~~,
\eeq
where $\GG$ is independent of $R$~. If we make this assumption then
then we can deduce the $R$-dependence of $\tt$ from the measurement of ${\cal G}_{SS}(R)$~.

It would be interesting if this scenario could be generalized to
a set of observables $A_i(R,T)$ $i=1,\ldots,N$ for which the 
correlators have the form
\beq
\la A_i(R,T)A_j(R,0)\ra=\sum_{\aa=\pm}\GG_{i\aa}(R)e^{-V_\aa(R)T}\GG_{j\aa}(R)~~,
\eeq
where the $R$-dependence of the coupling coefficients comes from the
movement of the mixing angle. That is
\beq
\GG_{i\aa}(R)=\sum_{a=S,M}\GG_{ia}O_{a\aa}(R)~~.
\eeq
Introducing the $N$-component vectors ${\bf C}_\aa=\{\GG_{i\aa}\}$ and ${\bf C}_a=\{\GG_{ia}\}$
we easily see from the the form of the orthogonal matrix $O$ given in eq(\ref{ORTHO})
that
\beq
{\bf C}_+(R)\wedge {\bf C}_-(R)={\bf C}_S\wedge {\bf C}_M~~,
\eeq
and
\beq
{\bf C}_{i+}(R)^2+{\bf C}_{i-}(R)^2={\bf C}_{iS}^2+{\bf C}_{iM}^2~~~~~~~~i=1,\ldots,N~~.
\eeq
If the left sides of these equations do show a lack of dependence on $R$ then we would
have a good test of the the above coupling hypothesis. It is not obvious that all sets of
operators have this property but it may be possible to construct a sufficient set.
However we might reasonably expect in general that the $R$-dependence of these
quantities is relatively weak compared to that induced by the rapid variation of the 
mixing angle through the mixing region.

On the assumption that an acceptable set of operators can be found, the 
dependence of $\tt$ on $R$ can be elucidated by examining the quantities 
$y_+={\bf n}.{\bf C}_+(R)$ and $y_-={\bf n}.{\bf C}_-(R)$ where ${\bf n}$ is any appropriate $N$-vector. 
They have the form
\beq
\left(\bea{c}y_+\\y_-\eea\right)=\left(\bea{cc}\cos\tt&-\sin\tt\\
                      \sin\tt&\cos\tt\eea\right)\left(\bea{c}y_S\\y_M\eea\right)~~,
\eeq
where $y_S={\bf n}.{\bf C}_S$ and $y_M={\bf n}.{\bf C}_M$~. Clearly the point $(y_+,y_-)$ lies on
a circle and $\tt$ is the angle, referred to an appropriate origin, that
fixes the position of the point on the circle. 

\section{Light Quarks with Spin}

Static quark calculations are particularly relevant to the spectator
quark approximation in which, except for determining multiplicities, the 
heavy quark spins play essentially no r\^ole. Of course in the physical case of
bottomonium, the small mass differences associated with
the hyper-fine structure of the $B$-mesons are crucial in determining
allowed decays. These effects are however essentially kinematic and will be 
taken into account at the appropriate point. In examining the basic phenomenon 
of string breaking we will temporarily omit these spins from consideration. 
The light quark spins however play a crucial part in both string breaking and decay dynamics. 

As discussed in section 2 our crude model provides a generalization of the scalar case
to that of light quarks with spin. It remains a two-channel mixing problem
because the light quark-anti-quark pair emerges in a triplet spin state with 
definite orientation. We can achieve this result by introducing a transition 
potential of the form
\beq
V(\RR)=\left(\bea{cc}V_{Q{\bar Q}}(R)&V_{MS}(\RR)\\
                     V^{\dag}_{MS}(\RR)&E_{MM}(R)\eea\right)~~,
\label{POT}
\eeq
where now the spin label structure of the transition element is given by
\beq
V_{MS\aa\bb}(\RR)=\fr{1}{\sqrt{2}}V_I(R)(\ss_i{\hat{\RR}}_i)_{\aa\bb}~~.
\eeq
This indeed implies that $V_{MS}(\RR)$ couples the
static quark-anti-quark state to a two-meson state in which the light quark
spins are in a state represented by the wave function $\xi_{\aa\bb}$ given
in eq(\ref{LQWF1})~.
Projecting down onto the appropriate subspace the two channel structure
emerges and the effective potential matrix is 
\beq
V^{\mbox{eff}}(R)=\left(\bea{cc}V_{Q{\bar Q}}(R)&V_{I}(R)\\
                     V_{I}(R)&E_{MM}(R)\eea\right)~~,
\label{EFF_POT}
\eeq
This is of precisely the form of the interaction matrix in the scalar case
and the eigenvalues $V_{\pm}(R)$ which which are given by eq(\ref{EIGV})
are the energies that will be computed on the lattice
from the measurement of appropriate operators. In the same spirit as for
the scalar calculation we can expect, through careful lattice measurements,
to determine the entries in $V^{\mbox{eff}}(R)$~. Having done that we revert
to considering the Schr\"odinger equation 
\beq
i\fr{\d}{\d t}\psi(\RR)=\left(\bea{cc}\fr{1}{2\mu_Q}\nab2&0\\
                        0&\fr{1}{2\mu_M}\nab2\eea\right)\psi(\RR)+V(\RR)\psi(\RR)~~,
\label{SE2}
\eeq
Following the pattern set out for the scalar case we calculate the decay rate 
by solving eq(\ref{SE2})  with $V_I(R)=0$ for an $S$-wave bound by $V_{Q{\bar Q}}(R)$ 
and a scattering wave controlled by $V_{MM}(R)$~. The transition rate is then
\beq
\GG=2\pi\sum_i\rho(k)|T_i(\bk)|^2~~,
\label{DECAYRATE}
\eeq
where 
\beq
T_i(\bk)=\la\bk~i|V_{MS}(\RR)|\psi_Q\ra~~,
\eeq
and $i$ is the polarization state of the triplet spin wavefunction.
In principle the state $|\bk~i\ra$ should be a solution of the the scattering problem
with the potential $V_{MM}(R)$. If we make the approximation of neglecting final state
interactions we can set
\beq
|\bk~i\ra=\fr{1}{\sqrt{2}}\ss_ie^{i\bk.\RR}~~.
\eeq
We have then
\beq
T_i(\bk)=\fr{1}{\sqrt{2}}\int d^3\RR e^{-i\bk.\RR}\mbox{\Tr}\{\ss_iV_{MS}(\RR)\}\psi_Q(R)~~.
\eeq
That is
\beq
T_i(\bk)=4\pi ik_i\int_0^{\infty}dRR
           \left(\fr{\cos kR}{k^2}-\fr{\sin kR}{k^3R}\right)V_I(R)\psi_Q(R)~~.
\label{GR3}
\eeq
Note that $T_i(\bk)\sim O(k)$ for small $k$~. This is the angular momentum barrier 
associated with the $P$-wave final state. 

\section{Heavy Meson Decay}

Currently there is no detailed information on string breaking in QCD at zero temperature
although there is a strong indication of the mixing effect at finite temperature \cite{DeT}.
The latest measurement of the $Q\bar Q$ potential with dynamical quarks does
not show an unequivocal mixing effect directly \cite{Burk}. This is not inconsistent
with the mixing scenario if the mixing range in $R$ and the split in eigenenergies
at maximal mixing is sufficiently small \cite{ITD1,ITD2,ITD3}. It is therefore 
interesting to reverse the above argument and use the information on the decay of 
the $\Upsilon(4S)$ to $B\bar B$ to estimate the energy split at maximum mixing that
we should expect to see on the lattice. The theory also enables us to  
make predictions for other bottomonium decays.

The hyperfine structure of the heavy-heavy and heavy-light quark systems has 
a strong effect on the relative positions of masses and thresholds and hence 
on decay rates. In order to make use of experimental results therefore, it is necessary 
to reinstate the spins of the heavy quarks. To compute the decay rates of real 
processes, we must take into account the spin structure of the open final state channels 
in order to assign the correct momenta to them.

The heavy quarks in the $\Upsilon$ are in a triplet state and continue, 
as spectator quarks, in that state after decay. We can represent this state by a spin
wave function $\ss_{j\aa'\bb'}/\sqrt{2}$, where $\aa'$ is the heavy quark label and
$\bb'$ is the heavy anti-quark label. In our model the light quarks are produced in a
triplet state that can be represented by a wave function 
$\ss_{i\aa\bb}/\sqrt{2}$~. A sufficiently complete basis for the achievable final states is
\beq
\psi_{ij}=\fr{1}{2}~\ss_{i\aa\bb}\ss_{j\aa'\bb'}~~.
\eeq
The transition matrix element we require is then
\beq
\la\psi_{i'j'}\bk|V_{MS}|\psi_Q,j\ra=\dd_{jj'}T_{i'}(\bk)~~,
\eeq
where $\bk$ is the relative momentum for the final state mesons and $T_{i'}(\bk)$
is the amplitude as previously defined. The spin state $\psi_{ij}$ can be decomposed into
a physical basis such as $B\bar B$, $B{\bar B}^*+B^*\bar B$, $B^*{\bar B}^*(S=2)$ and 
$B^*{\bar B}^*(S=0)$~. The resulting decay rate is given by eq(\ref{DECAYRATE})
modified by a probability factor $p$ appropriate to the channel under consideration.
\beq
\GG=p2\pi\rho(k)\sum_i|T_i(\bk)|^2~~,
\label{DECAYRATE2}
\eeq
These probabilities, calculated from the appropriate overlap coefficients in the 
spin wavefunction recoupling scheme, for the physical channels are
\n\n
\begin{center}
\begin{tabular}{|c|c|c|c|}\hline
\rw $B\bar B$&$B{\bar B}^*+B^*\bar B$&$B^*{\bar B}^*(S=2)$&$B^*{\bar B}^*(S=0)$\\\hline
\rw 1/12&1/3&5/9&1/36\\\hline
\end{tabular}
\end{center}
\n\n
Introducing the radial wavefunction $\chi(R)=\sqrt{4\pi}R\psi_Q(R)$ and using eq(\ref{TRANSPOT})
we can compute $T_i(\bk)$ from eq(\ref{GR3}) as
\beq
T_i(\bk)=\sqrt{4\pi} ik_iA(k,m_q)W~~,
\eeq
where
\beq
A(k,m_q)=\int_0^{\infty}dR
           \left(\fr{\cos kR}{k^2}-\fr{\sin kR}{k^3R}\right)e^{-m_qR}\chi(R)~~.
\eeq
Finally the decay rate to a particular channel is
\beq
\GG=pN_f4\mu k^3|A(k,m_q)|^2W^2,
\label{DECAYRATE3}
\eeq
where we have used $\rho(k)=\mu k/(2\pi^2)$, $\mu$ is the reduced mass
of the final state particles and $N_f$ is the number of degenerate $B\bar B$ channels
which is identical to the number of degenerate light quarks.
We use mass values from the Review of Particle Physics \cite{PARDAT},
\n\n
\begin{center}
\begin{tabular}{|r|r|r|r|r|r|r|}\hline
\rw Mass&$M_B$&$M_{B^*}$&$M_{B_s}$&$M_{B^*_s}$&$M_{\Upsilon(4S)}$&$M_{\Upsilon(5S)}$\\\hline
\rw GeV&5.279&5.325&5.369&5.416&10.580&10.860\\\hline
\end{tabular}\n\n 
\end{center}
We choose as a representation of $V_{Q\bar Q}(R)$ its parametrization in eq(\ref{CORNELL}). 
A choice of parameter values that accord reasonably well with lattice measurements and
the $\Upsilon(S)$ excitation spectrum is $\aa=0.52$ and $\ss=(0.429)^2$ GeV$~^2$, 
together with a heavy quark reduced mass $\mu_Q=2.1$ GeV. The resulting radial
wavefunctions $\chi(R)$ for $\Upsilon(4S)$ 
and $\Upsilon(5S)$ are shown in Fig 2. Both wavefunctions remain substantial over a range of 
$10~\mbox{GeV}^{-1}\simeq 2~\mbox{fm}$ with oscillations near the origin of a half
wavelength $\sim~1~-~2~\mbox{GeV}^{-1}$ before reaching broad maxima at 
$\sim~5~\&~7~\mbox{GeV}^{-1}$~.

As indicated above we ignore final state interactions, effectively setting $V_{MM}(R)=0$~.
Because our calculation is a preliminary one we will assume in what follows that $m_q=0$~. 
This may be justified as a rough approximation by the low masses of dynamical quarks. 
A crucial factor in computing the decay rate is the amplitude $A(k,m_q)$. We exhibit this
amplitude, for the case $m_q=0$, as a function of the final state momentum $k$ in Fig 3
for the $\Upsilon(4S)$ and $\Upsilon(5S)$. Note that in both cases it is small in 
magnitude for $k>0.5$ GeV. We can expect therefore as a general rule that only decays
with a sufficiently small final state momentum will contribute substantially to the
width of the initial state.

In our model the only open channel for $\Upsilon(4S)$-decay is $B\bar B$~. Experimentally
this process dominates. The relative momentum of the the $B\bar B$ mesons in the final 
state is $k=\sqrt{M_B(M_{4S}-2M_B)}=.3408$GeV~. This yields a value for the
amplitude $A(k,0)=-10.4$~. From eq(\ref{DECAYRATE3}) with $N_f=2$ and $p=1/12$ we find
\beq
\GG(\Upsilon\rightarrow B \bar B)= 7.02~W^2~~.
\eeq
The experimental result is $\GG(\Upsilon\rightarrow B \bar B)=0.010(5)$ GeV. This yields
an estimate $W=0.038(9)$ Gev. In turn we can estimate the eigenenergy split at maximum
mixing as $\Delta E=2\sqrt{N_f}W=0.11(3)$ GeV~. We can crudely estimate the range in $R$
over which the mixing takes place as
\beq
\Delta R=\fr{\Delta E}{V'_{QQ}(R_c)}\simeq\fr{\Delta E}{\ss}~~,
\eeq
with the result $\Delta R\simeq 0.6~\mbox{GeV}^{-1}\simeq 0.12$fm~.
To resolve such a split requires a rather fine lattice. The small value 
of this split, coupled with the decoupling of the two-meson state outside the
mixing region may be regarded as an explanation of why current measurements of 
the Wilson loop do not yet reveal string breaking in lattice QCD.

We can extend the above analysis to higher excitations such as the $\Upsilon(5S)$
which has a number of open decay channels. The results, assuming $W$ has the same value
as before, are
\n\n
\begin{center}
\begin{tabular}{|c|c|c|c|c|}\hline
Channel&$N_f$&k(GeV)&$A(k,0)$&$\GG$(MeV)\\\hline
\rw $B\bar B$&2&1.26&-0.68&2.0\\
\rw $B{\bar B}^*+B^*\bar B$&2&1.16&-0.734&9.0\\
\rw$B^*{\bar B}^*$&2&1.06&-0.430&4.0\\
\rw$B_s{\bar B}_s$&1&0.81&1.651&2.0\\
\rw$B_s{\bar B}_s^*+B_s^*\bar B_s$&1&0.636&1.l545&3.0\\
\rw$B_s^*{\bar B}_s^*$&1&0.389&-7.33&29.0\\\hline
\end{tabular}
\end{center}
\n\n
The total width is $\GG\simeq 50$ MeV which is to be compared with the
the quoted experimental result $\GG=110(13)$ MeV. Given the crudity of the
theoretical approach and the experimental difficulties this is not an unreasonable
comparison. 

\section{Conclusions}

We have examined the relevance of string breaking to the decay of 
heavy quark systems. The anaysis shows how string breaking on the lattice 
can be related to the decay rates for such processes as 
$\Upsilon(4S)\rightarrow B{\bar B}$~. The central idea is that the
string breaking can be viewed as a two channel process. This is
controlled by a $2\times 2$ potential matrix of which the diagonal elements
represent the interquark potential and the energy and interaction potential
of two static mesons. The off-diagonal elements correspond to a transition
potential between the two channels. Just as the interquark potential can be used
in a dynamical calculation of heavy quark states so the transition potential
can be used to compute the transition rate from the bound quark system to
two freely moving heavy mesons. This is a standard relatively
non-controversial quantum mechanics calculation. The main assumption behind it is 
that the Born-Oppenheimer approximation that underpins the heavy quark calculation
holds good when the transition matrix elements are included. This is 
equivalent to assuming that there is no strong retardation effect in the 
transmission of the dynamical quark. If this quark mass is sufficiently
low this assumption is not unreasonable.

At the moment there are no measurements in QCD that permit an accurate 
realization of the scheme. However, with some simplifying assumptions  
on the form of the elements of the potential matrix, it is possible to make a preliminary
calculation of the decay rate of the $\Upsilon(4S)$ in terms of the 
energy parameter $W$, yielding a value $\simeq 38$ MeV. This gives rise to   
an eigenenergy split of $\sim 110$ MeV. Although this suggests that the split
may be hard to detect on the lattice the study of mixing remains an important goal 
because of its physical significance. Our analysis also strongly supports
the desirability of measuring an appropriate suite of operators for the purpose
of measuring the movement of the mixing angle \cite{PhWi,KnSo}. In fact, as 
indicated in \cite{ITD3}, it would be very useful in estimating $V_I(R)$ to
measure the transition amplitude for string to two mesons in the quenched
appoximation \cite{StKo}.

Taking the above value of $W$ as a guide we are able to use the formalism to
compute the partial widths for $\Upsilon(5S)$-decay.  We find that the dominant
decay is through the $B^*_s{\bar B}^*_s$ channel. Our computed total width is
roughly half the measured width. We feel that this result is not unreasonable
given the preliminary character of the calculation and the limited information
available from experiment on decay channels.

We conclude that with some improvement of lattice measurements it will be possible to 
confront string breaking with experimental results when these become available. 

\pagebreak

\newpage
{~~~~~~~~~~~~~~~~~~~~~~~~~~~~~~~~~~~~~~~~~~}
\vskip 20 truemm
\begin{figure}[htb]
\begin{center}\leavevmode
\epsfxsize=10 truecm\epsfbox{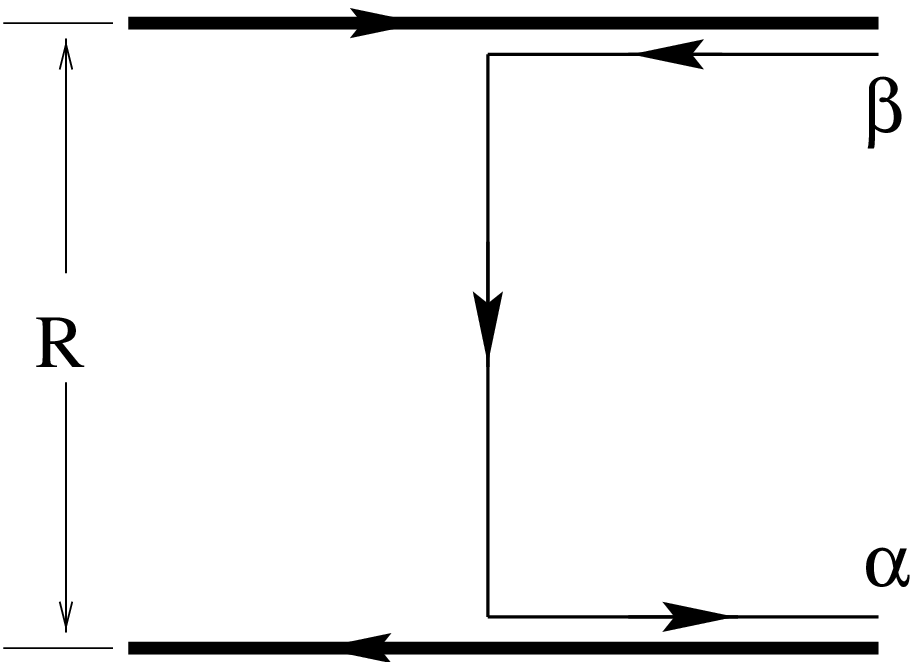}
\end{center}
\caption[]{Transition from string to two-meson state. Heavy lines
represent static quarks, light lines represent dynamical quarks.}
\label{figure:SBFIG}
\end{figure}
\vskip 10 truemm
\begin{figure}[htb]
\begin{center}\leavevmode
\epsfxsize=15 truecm
\epsfbox{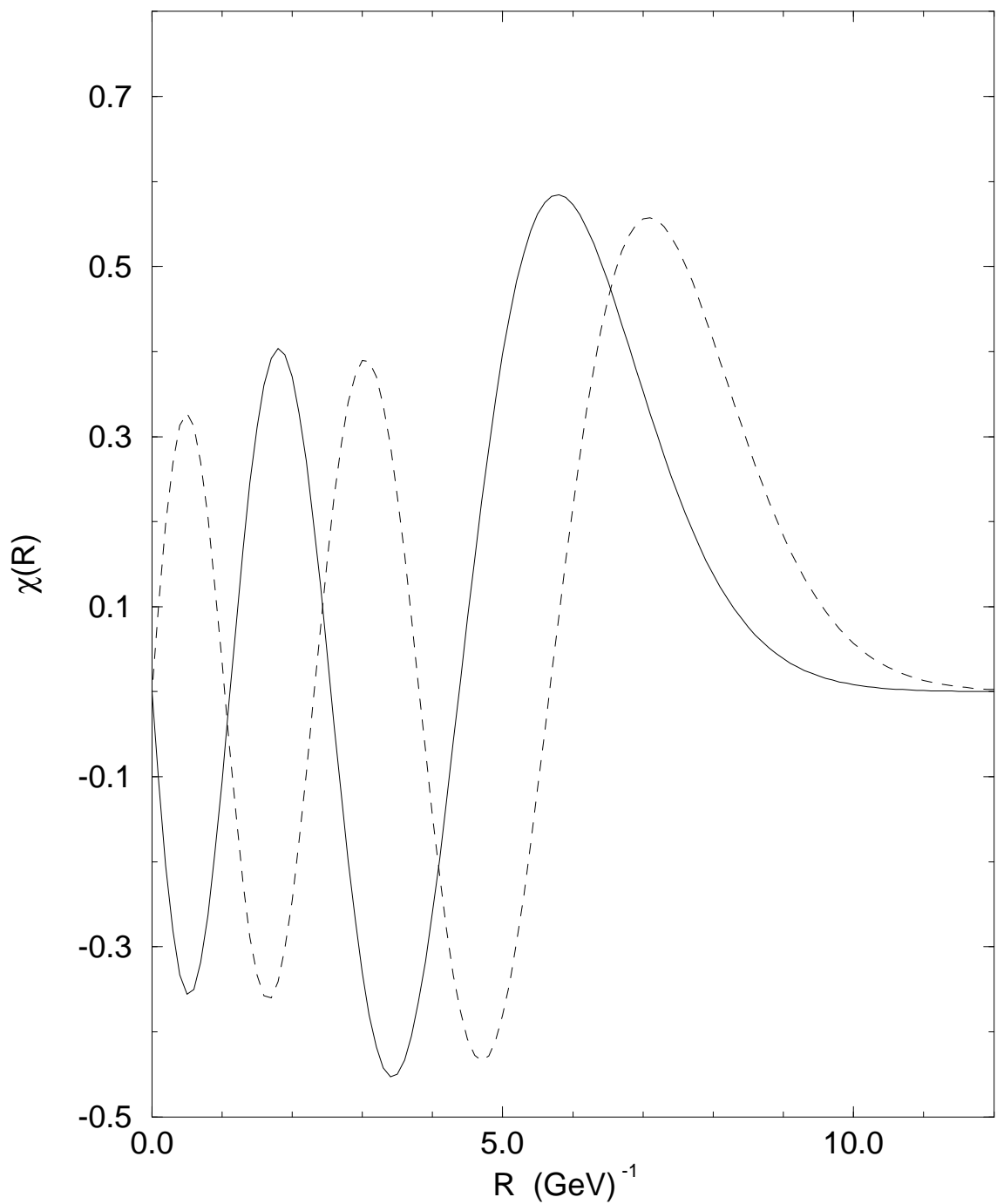}
\end{center}
\caption[]{Wavefunctions for $\Upsilon(4S)$ and $\Upsilon(5S)$.}
\label{figure:UPSILON}
\end{figure}
\vskip 10 truemm
\begin{figure}[htb]
\begin{center}\leavevmode
\epsfxsize=15 truecm
\epsfbox{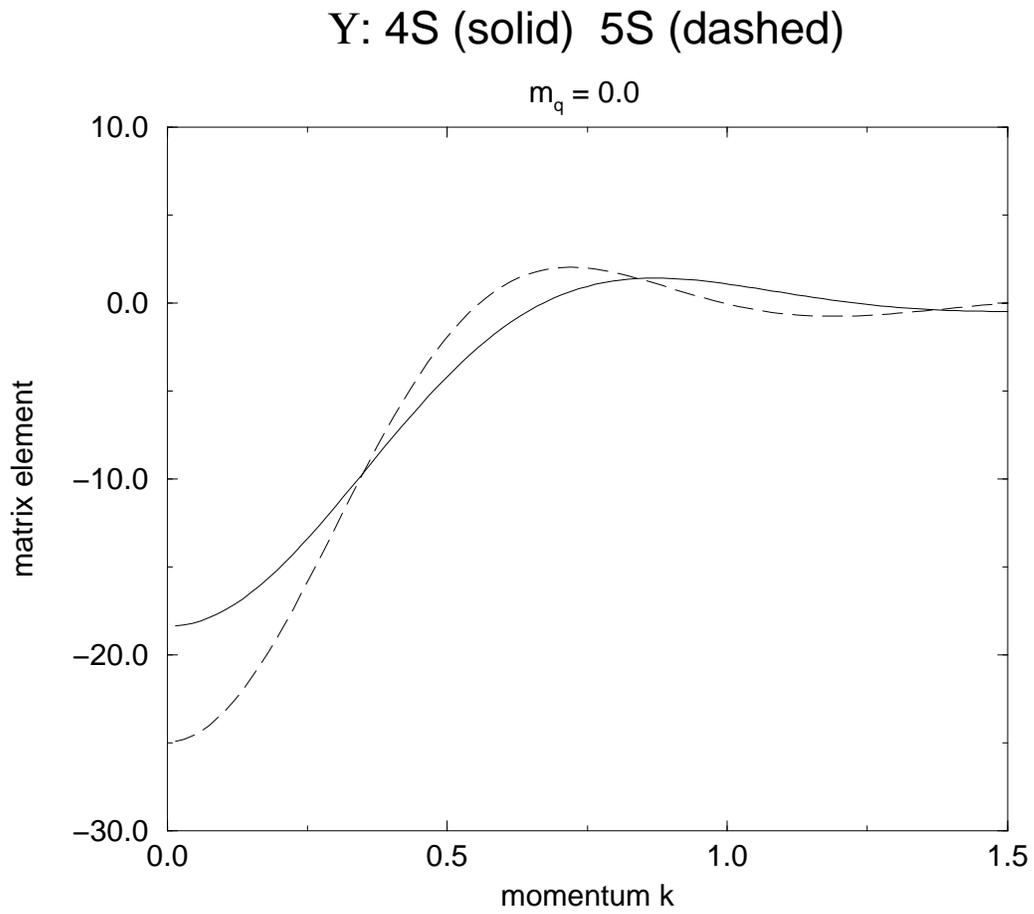}
\end{center}
\caption[]{Transition matrix element, $A(k,0)$, for $\Upsilon(4S)$ and $\Upsilon(5S)$.}
\label{figure:AMP}
\end{figure}

\end{document}